\documentclass[12pt]{spieman}  
\usepackage{amsmath,amsfonts,amssymb}
\usepackage{graphicx}
\usepackage{setspace}
\usepackage{tocloft}
\usepackage{adjustbox}
\usepackage{caption}
\usepackage{subcaption}
\usepackage{booktabs}
\usepackage{siunitx}
\usepackage{multirow}
\usepackage{tikz}

\title{Standardized CycleGAN training for unsupervised stain adaptation in invasive carcinoma classification for breast histopathology}

\author[a]{Nicolas Nerrienet}
\author[a]{Rémy Peyret}
\author[a]{Marie Sockeel}
\author[a]{Stéphane Sockeel}
\affil[a]{Primaa, 41 rue du Sentier, Paris, France, 75002}

\cftpagenumbersoff{figure}
\cftpagenumbersoff{table} 
\begin{document} 
\maketitle

\begin{abstract} 

\textbf{Purpose}: Generalization is one of the main challenges of computational pathology. Slide preparation heterogeneity and the diversity of scanners lead to poor model performance when used on data from medical centers not seen during training. In order to achieve stain invariance in breast invasive carcinoma patch classification, we implement a stain translation strategy using cycleGANs for unsupervised image-to-image translation.
Those models often suffer from lack of proper metrics to monitor and stop the training at a particular point. In this work, we also introduce a novel method to solve this issue.

\textbf{Approach}: We compare three cycleGAN-based approaches to a baseline classification model obtained without any stain invariance strategy. 
Two of the proposed approaches use cycleGAN's translations at inference or training in order to build stain-specific classification models. 
The last method uses them for stain data augmentation during training. This constrains the classification model to learn stain-invariant features.
Regarding CycleGANs' training monitoring, we leverage Fréchet Inception Distance (FID) between generated and real samples and use it as a stopping criterion.
We compare CycleGANs' models stopped using this criterion and models stopped at a fixed number of epochs.

\textbf{Results}: Baseline metrics are set by training and testing the baseline classification model on a reference stain. 
We assessed performances using three medical centers with H\&E and H\&E\&S staining. 
Every approach tested in this study improves baseline metrics without needing labels on target stains. 
The stain augmentation-based approach produced the best results on every stain. Each method's pros and cons are studied and discussed in this paper.
Moreover, FID stopping criterion proves superiority to methods using a predefined number of training epoch and has the benefit of not requiring any visual inspection of cycleGAN results.

\textbf{Conclusion}: In this work, we introduce a method to attain stain invariance for breast invasive carcinoma classification by leveraging CycleGANs abilities to produce realistic translations between various stains. Moreover, we propose a novel systematical method for scheduling cycleGANs' trainings by using Fréchet Inception Distance (FID) as a stopping criterion and prove its superiority to other methods. Finally, we give an insight on the minimal amount of data required for cycleGAN training in a digital histopathology setting.

\end{abstract}


{\noindent \footnotesize\textbf{*}Nicolas Nerrienet,  \linkable{nicolas.n@primaalab.com} }

\begin{spacing}{2}   

\section{Introduction} \label{sec : Introduction}
\textbf{General context}. Digital histopathology is changing the way pathologists work on a day-to-day basis. Not only does it allow them to share data and insights about a diagnosis across the globe easily, but with computational pathology, algorithms capabilities may be leveraged to ease pathologists' work on many tasks. Deep learning has shown promising results for various applications such as cancer classification \cite{breast_cancer_classification, breast_deep_learning, digital_pathology_deep_learning_survey}, detection \cite{weakly_supervised_glomeruli_detection_segmentation, nephritis_classification_detection, mitosis_detection} or segmentation \cite{nuclei_segmentation, gland_segmentation} of biomarkers required for diagnosis.

However, the integration of AI in histopathology comes with a set of challenges to overcome.
In particular, the diversity and heterogeneity in slide preparation increases data variability across centers or laboratories.
Slide preparation is a long and tedious process consisting of multiple steps which can each be a source of variability that negatively impacts algorithms performance \cite{impact_staining_scanner_variation}.
Specifically, biopsy samples are collected during surgeries and are molded into a paraffin bloc for easier slicing. In order for the paraffin to penetrate the tissue down to molecular level, the tissue is gradually dehydrated using series of ethanol baths with rising concentrations. Dehydration may affect tissue structure depending on the exact protocol applied. Samples are then sliced into thin tissue pieces and placed on a glass slide. As tissue handling is tedious, artefacts such as rips or folds often appear.
Additionally, the tissue being originally transparent, contrasting agents are to be added in order to analyse certain structures of interest.

This process is called staining and is a crucial part of slides preparation.
Hematoxylin and Eosin (H\&E) are typically used to dye cell nuclei in a specific color and tissue surrounding cells in another one, whereas some other stains, based on Immunohistochemical (IHC) reactions, target some specific receptors or molecular compounds of the cell membrane, making them visible. The staining process and the manufacturer from whom the staining agents are purchased differ from a laboratory to another, leading to intra-stain variability. In France, pathologists are also used to adding Safran to H\&E (H\&E\&S) in order to dye connective tissues. This gives yet another appearance to tissue samples, which can greatly differ depending on the balance of Hematoxylin's and Eosin's constitutive components.
Digital pathology adds one last source of variability. Physical slides are scanned to produce Whole Slide Image (WSI) that can be viewed on a software. The scanner's brand and settings, as well as the file format, are clearly identified sources of data variability \cite{impact_staining_scanner_variation}.
These considerations result in WSI having clearly distinguishable appearances from one laboratory to another. 

In fact, it has been observed that AI models performed poorly when applied to data coming from medical centers not seen during training or to data stained differently than training data \cite{impact_staining_scanner_variation, tellez_quantify_augmentation_normalization}. Therefore, before being used in a new laboratory, models need to be fine-tuned using annotated data from this laboratory. This is a time-consuming and expensive process as it requires expert pathologists.

Hence, generalization is one of the main challenges in computational histopathology and solving it would allow a unique model to be applied on a wide range of laboratories without fine-tuning or annotated data \cite{embracing_imperfect_dataset}.

\textbf{Related works}. 
This issue is commonly addressed in literature using two global approaches: normalization and augmentation.
Most of the time, normalization is used to reduce dissimilarity within a given domain \cite{importance_stain_normalization, adaptive_color_deconvolution, macenko}. For instance, the
simplest form of stain normalization is to simply convert RGB images to grayscale, assuming that most of
the H\&E signal is contained in the morphological and structural patterns of the tissue.

On the other hand, color augmentation is often about synthesizing new images by varying the colors, the contrast or the illumination. This is used in a deep learning setting, to make learnt features independent of color variations. 

Recently, due to their ability to generate highly realistic outputs, Generative Adversarial Networks (GAN) \cite{GAN} and their derivatives have been also intensively used to address the generalization challenge in digital histopathology.
Specifically, either Pix2Pix \cite{pix2pix} or CycleGANs \cite{CycleGAN} are utilized as stain translation devices depending on the need or availability of paired samples. 
The Pix2pix often displays more realistic and consistent results but requires paired samples.
However, in a digital histopathology context, it can be problematic to obtain paired datasets, as it involves the ability to re-stain stained samples which is a costly and tedious process.
Consequently, most studies using GANs’ setup in digital histopathology rely more heavily on Cycle-GANs. \cite{CycleGAN} demonstrated that cycleGANs can produce samples in a highly realistic fashion without the need for paired training data.

Vasilijesic et al\cite{towards} and Gadermayr et al\cite{gadermayr_MDS} identified three methods that can be used to address domain adaptation issues with CycleGANs, namely Multi-Domain Supervised 1 and 2 (MDS1/MDS2) and Unsupervised Domain Augmentation (UDA). For each CycleGAN based method, a deep learning task is used on top of the CycleGAN’s translations. In these approaches, the source domain is either the domain on which most of the labeled data is available or the domain on which models such as classifiers or detectors already perform well. Conversely, the target domain is often scarce in annotated data and exhibits poor model performance, or at least, poor generalization performance. In short, MDS1 consists of training a classifier on the source domain and applying it on target domain samples translated to the source domain, whereas MDS2 aims at training a classifier on source domain samples translated to the target domain. Finally, UDA’s goal is to use CycleGANs’ translation as a mean of data augmentation during training \cite{towards}.

In the literature, Shaban et al\cite{stainGAN} employs CycleGANs to show its superiority compared to conventional methods in the case of lymph node metastasis classification. However, they solely use the MDS1 method and lack the exploration of the other ones which were proven to lead to better performing models\cite{towards}. Moreover, little insight is given on how the final CycleGAN model was chosen and on the data it was trained on. Several other studies use these models in the case of virtual staining for mitosis detection \cite{Mitosis_midog_domain_generalization, mitosisCycleGAN} or in the case of glomeruli segmentation \cite{towards} among other works. Interestingly, Vasilijesic et al\cite{towards} points out a weakness in using CycleGANs, that is, the visual quality of a CycleGAN translation is not always related to the performance of the task on top of the generation process. This unexpected behavior implies a substantial difficulty not only when comparing and choosing CycleGANs for the task at hand, but also when one must decide to stop the training process or not.

This study is built upon all this prior knowledge and applies CycleGANs for breast tumor classification generalization, while introducing a consistent method on how to evaluate and choose image-to-image translation models for the downstream task and thus, addressing the previously mentioned concern.

\textbf{Contributions}. This study makes the following key contributions:
\begin{itemize}
    \item We introduce a systematic and consistent method to train and choose a CycleGAN model.
    \item We study the optimal amount of WSI required to train a CycleGAN for breast tumor classification.
    \item We introduce a stain robust breast tumor classification model.
    \item We prove the relevance of using CycleGANs to attain stain invariance in invasive breast carcinma classification tasks on a novel large multi center dataset.
\end{itemize}

In the following section, a description of deep learning models utilized in this study is exposed, as well as the datasets used. Section 3 describes the different experiments performed in this work, section 4 presents the experimental results.
In the final section, we will present and interpret the results obtained from our research.

\section{Materials and Methods} \label{sec : material & methods}

\subsection{Datasets} \label{subsec : material & methods : datasets}
\begin{figure}[htp]
    \centering
    \begin{subfigure}[b]{0.23\textwidth}
        \centering
        \includegraphics[width=\textwidth]{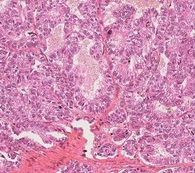}
        \caption{Source center}
        \label{fig:source_center}
    \end{subfigure}
    \begin{subfigure}[b]{0.23\textwidth}
        \centering
        \includegraphics[width=\textwidth]{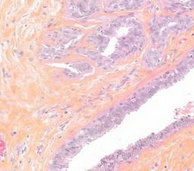}
        \caption{Center 1}
        \label{fig:center_1}
    \end{subfigure}
    \begin{subfigure}[b]{0.202\textwidth}
        \centering
        \includegraphics[width=\textwidth]{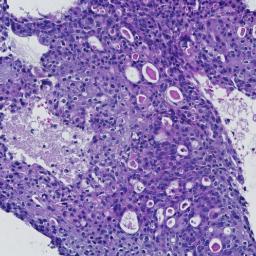}
        \caption{Center 2}
        \label{fig:center_2}
    \end{subfigure}
    \begin{subfigure}[b]{0.202\textwidth}
        \centering
        \includegraphics[width=\textwidth]{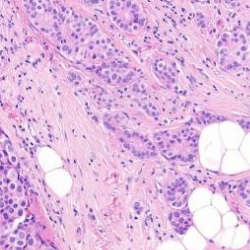}
        \caption{Center 3}
        \label{fig:center_3}
    \end{subfigure}
    \caption{Samples from various centers and their respective stains}
    \label{fig:different_stains_centers}
\end{figure}

The data used in this study comes from four different centers. Samples are collected from breast cancer patients that underwent breast biopsy or mastectomy. 
Slides were annotated through a two-phase process. Pathologists were asked to circle regions of the slides that contained IC (Invasive Carcinoma) and regions that contained CIS (Carcinoma In-Situ) or benign lesions as well as normal tissue structures. These annotations were further reviewed and validated by experienced pathologists. IC regions were categorized under "IC" while the non-IC regions were grouped into a "REST" category. To build the patches datasets, labeled regions were divided into patches of 256*256 pixels at magnification x5.
The reference center is the one with the largest labeled dataset. This center is here referred to as reference center or source center, interchangeably.
Source center, centers 1, 2 and 3 respectively used Hamamatsu S360 (X40), 3DHistech P1000 (X40), Apeiro Leica GS2 (X40) and 3DHistech P250 (X40) scanners. Samples from center 1 are stained with H\&E\&S, while samples from center 2 and 3 are stained with H\&E.
Figure \ref{fig:different_stains_centers} depicts a sample from each of these centers, illustrating the high variability of observable appearances. 

For the specific case of the source center, the data is partitioned in training and testing sets at the case level, ensuring that slides from one case are exclusively present either in the train set or the test set. Consequently, train and test set cannot contain patches extracted from the same slide. The training set has been balanced with an equal number of 'IC' and 'REST' patches. For test sets however, labels distribution is unbalanced, and kept as such to reflect clinical reality. The amount of slides, patches from every center and for each dataset, as well as the label distribution and scanners are illustrated in the following Table \ref{tab:slides_patches_distribution}.

\begin{table}[h]
\centering
\resizebox{\textwidth}{!}{%
\begin{tabular}{cccccc}
\toprule
& \multicolumn{5}{c}{Centers} \\
\cmidrule(lr){3-6}
& & Source center & Center 1 & Center 2 & Center 3 \\
\midrule
\multirow{2}{*}{Slides} & train & 1690 & - & - & - \\
& test & 309 & 42 & 119 & 44 \\
\midrule
\multirow{2}{*}{Patches} & train & 1.7M & - & - & - \\
& test & 469,217 & 75,888 & 35,209 & 46,808 \\
\midrule
\multirow{2}{*}{Label distribution (test sets)} & IC & 205,380 & 19,229 & 10,879 & 33,145 \\
& REST  & 263,837 & 56,659 & 24,330 & 13,663 \\
\midrule
\multirow{2}{*}{Scanners} & scanner & Hamamatsu S360 & 3DHistech P1000 & Apeiro Leica GS2 & 3DHistech P250 \\
& magnification & \multicolumn{4}{c}{X40} \\
\bottomrule
\end{tabular}
}
\caption{\label{tab:slides_patches_distribution}Distribution of slides, patches, label distribution, and scanners across centers.}
\end{table}

\subsection{Methods} \label{subsec : methods}

\subsubsection{Translation model} \label{subsubsec : material & method : methods : translation model}
\begin{figure}[htp]
    \centering
    \includegraphics[width=1\textwidth]{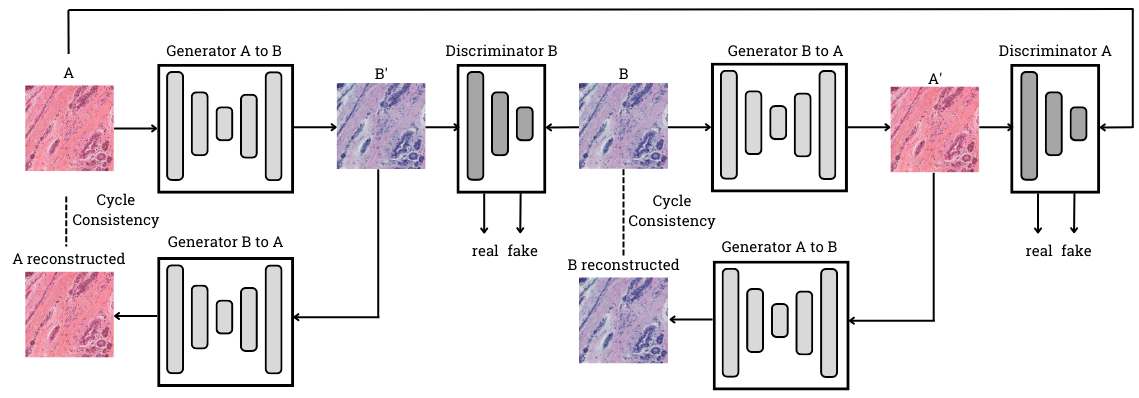}
    \caption{CycleGAN architecture. Generator $A \longleftrightarrow B$ is responsible for taking a sample from a domain $A$ and translate it into domain $B$, while the discriminator $B$ classify the translated samples as fake or a true sample from the domain $B$. Translation unicity is guaranteed by minimizing the difference between the original sample and the translated sample reconstructed in domain $A$ by generator $B \longleftrightarrow A$. The same logic is applied for the other generator.}
    \label{fig:cyclegan_global}
\end{figure}

In this work, CycleGANs \cite{CycleGAN} (see Figure \ref{fig:cyclegan_global}), are used as translators between different stains prior to the breast carcinoma classification task.

\begin{figure}[htp]
    \centering
    \begin{subfigure}[b]{0.178\textwidth}
        \centering
        \includegraphics[width=0.75\textwidth]{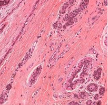}
        \caption{Source center}
        \label{MDS1:training}
    \end{subfigure}
    \begin{subfigure}[b]{0.173\textwidth}
        \centering
        \includegraphics[width=0.75\textwidth]{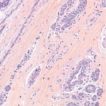}
        \caption{Center 1}
        \label{MDS1:inference}
    \end{subfigure}
    \begin{subfigure}[b]{0.175\textwidth}
        \centering
        \includegraphics[width=0.75\textwidth]{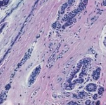}
        \caption{Center 2}
        \label{MDS1:inference}
    \end{subfigure}
    \begin{subfigure}[b]{0.175\textwidth}
        \centering
        \includegraphics[width=0.75\textwidth]{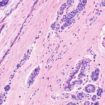}
        \caption{Center 3}
        \label{MDS1:inference}
    \end{subfigure}
        \caption{CycleGANs transformations examples from source to targets}
    \label{fig:cyclegan_transfo_examples}
\end{figure}

Figure \ref{fig:cyclegan_transfo_examples} shows an example of CycleGAN translations with our data. 
In this example, a patch from the source domain is translated to the other centers domain.
A CycleGAN is trained for every pair of centers (source center, center $i$) with $i \in \{1,2,3\}$, as the goal is to make translation between the reference center and every other center. 

\textbf{CycleGAN architecture}. CycleGANs' generators and discriminators architecture are described in the original paper \cite{CycleGAN}.

Johnson et al\cite{perceptual_losses_real_time_style_transfer} show that this architecture produces high quality outputs in the case of style transfer.
Moreover, they demonstrated that instance normalization greatly improves style-transferred samples quality as well as being more efficient for small batch size than batch normalization \cite{perceptual_losses_real_time_style_transfer, instance_normalization_missing_ingredient}. We thus adopt this normalization strategy.
As problems such as vanishing gradients, mode collapse and failing to converge are quite standard issues with GANs \cite{gans_challenges&solutions}, least-square-loss is used instead of the negative-log-likelihood standard loss to ensure a more stable training following \cite{lsGAN}.

\textbf{CycleGAN training and evaluation}. The training of CycleGANs involves a specific number of patches, as determined by an experiment that is further elaborated upon in Section \ref{subsec : experiments : slides_number}.
The number of WSI from which we extract these patches is center-dependent as established by an experiment detailed within the same section.

Besides, the difficulty of evaluating generative models is a known issue.
Initially, generative models were evaluated by mere visual inspection. However, in digital histopathology, it would require expert knowledge to ensure that generated samples are realistic enough. In addition, it has been shown that generated samples' visual quality is not always correlated with performance for the task on top of the generation process\cite{towards}. Specifically, two CycleGANs from two different epochs with the same visual generation quality can yield significantly different results in the case of a segmentation task on the generated samples.
Hence, there is a need to evaluate the quality of a cycleGAN so as to monitor and stop its training in a releavant way. To do so, we propose to use Fréchet Inception Distance (FID) \cite{FID_metric} which is a popular metric used for generative models' evaluation. 
This metric uses an Inception model to extract features from generated and real images, those feature distributions are then used to compute the Fréchet Distance \cite{Frechet}.

\textbf{CycleGAN methods}. CycleGANs can be utilized to address generalization challenges in three distinct ways, each with its own unique characteristics and applications.

\begin{figure}[htp]
    \centering
    \begin{tabular}{c|c}
        \begin{subfigure}[b]{0.34\textwidth}
            \centering
            \caption{Training}
            \includegraphics[width=1\textwidth]{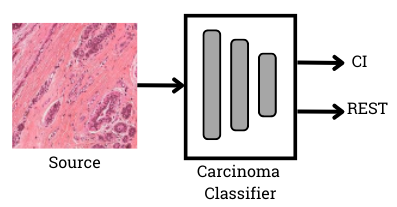}
            \label{MDS1:training}
        \end{subfigure}
        &
        \begin{subfigure}[b]{0.65\textwidth}
            \caption{Inference}
            \centering
            \includegraphics[width=1\textwidth]{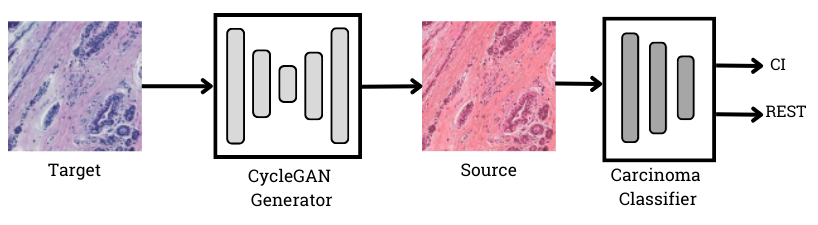}
            \label{MDS1:inference}
        \end{subfigure}
    \end{tabular}
    \vspace{5pt}
    \caption{Explains the Multi-Domain Supervised 1 (MDS1) process. \ref{MDS1:training} represents the training phase of MDS1 : the carcinoma classifier is trained on "real" source samples. \ref{MDS1:inference} represents the inference phase : the carcinoma classifier is used at inference time on "fake" target samples.}
    \label{fig:MDS1}
\end{figure}

In MDS1 \cite{gadermayr_MDS}, the CycleGAN's translation is used as a normalization preprocessing step to the main classification model at inference time. Samples are translated to the source domain before being classified. This enables the use of pre-trained source domain-specific classification models. This method is illustrated in Figure \ref{fig:MDS1}.

\begin{figure}[htp]
    \centering
    \begin{tabular}{c|c}
        \begin{subfigure}[b]{0.64\textwidth}
            \centering
            \caption{Training}
            \includegraphics[width=1\textwidth]{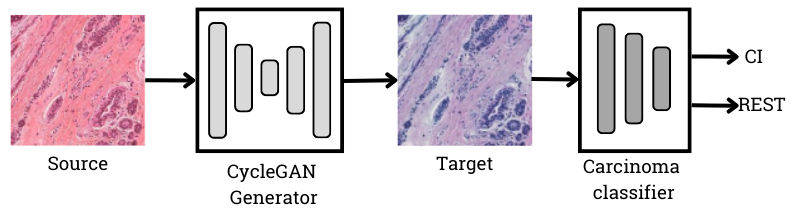}
            \label{MDS2:training}
        \end{subfigure}
        &
        \begin{subfigure}[b]{0.35\textwidth}
            \caption{Inference}
            \centering
            \includegraphics[width=1\textwidth]{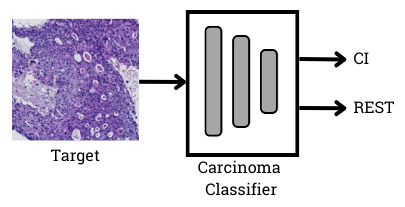}
            \label{MDS2:inference}
    \end{subfigure}
    \end{tabular}
    \vspace{5pt}
    \caption{Explains the Multi-Domain Supervised 2 (MDS2) process. \ref{MDS2:training} represent the training phase of MDS2 : the carcinoma classifier is trained on "fake" target samples. \ref{MDS2:inference} represent the inference phase : the carcinoma classifier is used at inference time on "real" target samples.}
    \label{fig:MDS2}
\end{figure}

In MDS2 \cite{gadermayr_MDS}, target domain-specific classification models are built using CycleGAN's translations as a normalization preprocessing step at training time as illustrated by Figure \ref{fig:MDS2}. Models built in this fashion can then directly be applied on target domain data. 

\begin{figure}[htp]
    \centering
    \begin{tabular}{c|c}
        \begin{subfigure}[b]{0.55\textwidth}
            \centering
            \caption{Training}
            \includegraphics[width=1\textwidth]{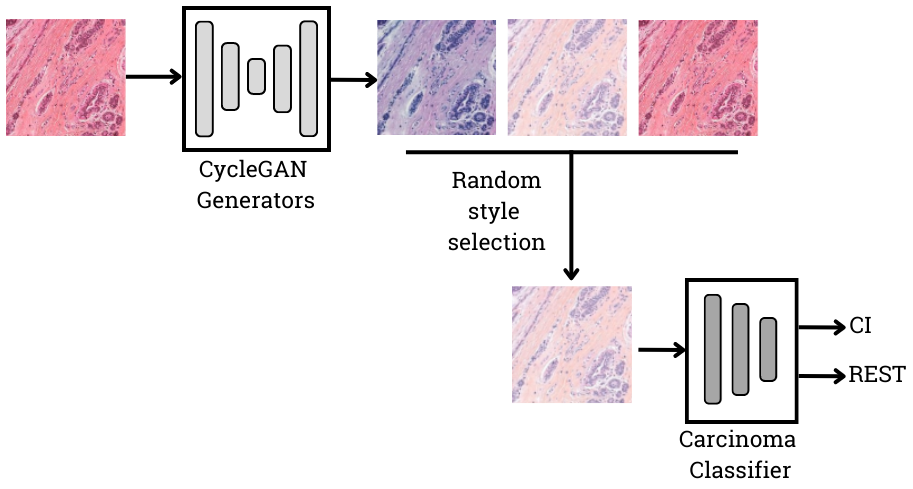}
            \label{UDA:training}
        \end{subfigure}
    &
        \begin{subfigure}[b]{0.44\textwidth}
            \caption{Inference}
            \centering
            \includegraphics[width=1\textwidth]{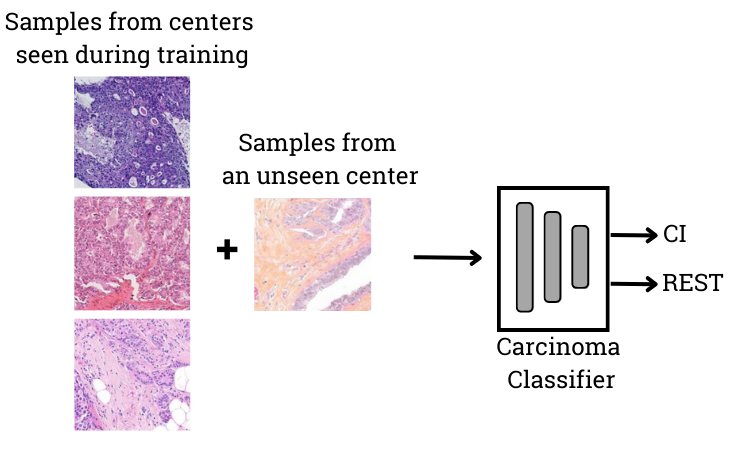}
            \label{UDA:inference}
        \end{subfigure}
    \end{tabular}
    \vspace{5pt}
    \caption{Explains Unsupervised Domain Augmentation (UDA) process. \ref{UDA:training} represents the training phase of UDA : "fake" samples from various CycleGANs translation are randomly selected and fed to the carcinoma classifier model during its training. \ref{UDA:inference} represents the inference phase : in this case, the carcinoma classifier can be used on real samples from any center, even samples from center not seen by the classifier during training.}
    \label{fig:UDA}
\end{figure}

Finally, Unsupervised Domain Augmentation (UDA) \cite{towards}, illustrated in figure \ref{fig:UDA},
makes use of CycleGANs’ translations to various domains as a mean of data augmentation during
training. This leads to the model extracting stain invariant features for the task at hand, provided that the pool of domains is diverse enough. Contrary to the other methods that use a single domain translation and build domain specific models, this approach aims at producing models that perform on a large variety of stains and thus attain stain invariance.

A major advantage of these three approaches is that they don't require any labeling in the domain in which samples are translated to, considerably reducing the time and money investment that labeling requires.

In this study, each of these approaches is used to solve the generalization challenge for breast invasive carcinoma classification following an experiment scheme described in the section \ref{subsec : experiments : methods comparison}.

\subsubsection{Classification model} \label{ subsubsec : material & methods : methods : classification model}
\textbf{Classifier architecture}. In order to classify patches into the two classes discussed in section \ref{subsec : material & methods : datasets}, a classification model was employed on top of the translation process as described in figures \ref{fig:MDS1}, \ref{fig:MDS2} and \ref{fig:UDA}. 
The model used for this task is an EfficientNetB1 \cite{TanEfficientNet2019} with two outputs. 
In order to train this model, we used a category cross-entropy loss \cite{CrossEntropyLoss} and an Adam optimizer \cite{AdamOptimizer}. 
Additionally, to enhance the training data pool, various data augmentation techniques were randomly applied at each training step:
\begin{itemize}
  \item \textbf{Flip}: Flip the patch horizontally.
  \item \textbf{Rotate}: Rotate the patch by a multiple of 90 degree angle.
  \item \textbf{Color Jitter}: Shift hue, saturation and brightness of the patch.
  \item \textbf{Noise}: Add Gaussian noise to the patch.
\end{itemize}

\textbf{Baseline classification model}. In order to establish a baseline performance to compare the different methods results, we train a classification model using the architecture aforementioned. 
This baseline classification model is trained on the reference center trainset (see section \ref{subsec : material & methods : datasets}) for 150 epochs, and baseline performance on each center is obtained by testing this model on every center's test dataset.

\subsubsection{CycleGANs comparison method and notations} \label{subsubsec : cyclegans comparison method}
Throughout this study, comparing various CycleGANs in the breast classification task context will be required.
To accomplish this, we introduce a method to evaluate task-wise CycleGAN's performance. 
First, we establish the following notation: a CycleGAN responsible for translations from reference center to center $i$, and vice-versa, will be mentioned as follows : CycleGAN\_$i$.
Next, in order to evaluate a CycleGAN's performance related to the classification task, we will use MDS1 modality. To elaborate, CycleGAN\_$i$ is utilized to translate samples from a dataset to the reference center. Then, the baseline classification model is tested on these translated samples to obtain metrics that will be used for comparison purposes.
We further refer to this method as "MDS1 comparison".

MDS1 modality is preferred for this task as baseline classification model can be used as an evaluator whereas labeled training data is too scarce to train "baseline models" for other centers in our settings.
One can argue that using MDS1 only is insufficient and indeed previous work showed that translation direction is an important factor \cite{towards, CycleGAN}.
CycleGANs can generate higher quality transformations depending on translation's direction and can even hide information in the translated samples \cite{towards, adversarial_self_defense_Bashkirova_2019, cycleGAN_stenography_master_Chu_2017}. The latter effect can have an unexpected impact on deep learning models.
However, this is often the case when one domain has more information than the other, which implies that a direction will suppress information and the other will restore what has been lost \cite{adversarial_self_defense_Bashkirova_2019, cycleGAN_stenography_master_Chu_2017}.
In our case, this effect is greatly mitigated as the domains, H\&E stains, hold the same information content.
Therefore, we will solely rely on MDS1 comparison throughout the study.

\section{Experiments} \label{experiments}

\subsection{CycleGANs training stopping criterion} \label{subsec : experiments : cyclegans training stopping criterion}
A common challenge when working with GANs is to know when the model has converged and when we should stop the training procedure.
Usually, mere visual inspection of generated samples' quality is key to determine when the training should end. 
However, this procedure comes with two major downsides. 
First, it implies saving the model at each training epoch in order to be able to retrieve the epoch weights that gave the most satisfying results, which is memory consuming and requires a manual intervention. 
Secondly, as stated in the section \ref{subsubsec : material & method : methods : translation model}, visual inspection is not sufficient as image quality does not systematically reflects performance for the task on top of the generation process.

In this experiment, we evaluate the effectiveness of using the FID metric as an early stopping criterion for training CycleGAN models. 
The FID stopping criterion operates in the following manner: CycleGAN models are trained until FID ceases to decrease for a fixed number of "patience" epochs, which was set to 15 in this experiment.
We compare a model trained using the FID criterion to models trained for 25, 50, 75 and 100 epochs. 
This comparison is done for CycleGANs\_$i$ (see section \ref{subsubsec : cyclegans comparison method}) where $i$ takes the following values $\{1,2,3\}$.

The FID for CycleGAN\_$i$ training is computed as the Fréchet distance between 1000 samples randomly selected in the source center, and 1000 samples translated from center $i$ to the the source center. Latter samples plays the role of the "generated images".

The resulting CycleGAN models are then evaluated on center $i$ test dataset using MDS1 comparison described in section \ref{subsubsec : cyclegans comparison method}.
For each center, we repeat this experiment three times to compute error margins and ensure results repeatability. 

\subsection{Slides Requirements} \label{subsec : experiments : slides_number}
First, we seek to find the optimal and minimal number of training patches required for effectively training a CycleGAN. To achieve this, we conduct experiments by training CycleGANs with sets of 100, 1000, and 10000 patches for each source-to-target center pair. We then compare their FID at the end of the training process, which uses the previously described stopping criterion.

Next, when it comes to digital histopathology, an under-discussed but nonetheless decisive matter is the number of slides required for training a CycleGAN that produces satisfying enough samples.
This point is crucial for the method convenience considering the difficulty of collecting a sufficient amount of WSIs to train deep learning algorithms.
Moreover, in the specific case of domain adaptation, it is convenient for methods’ presented in this study to use the minimum amount of data as possible. Therefore, we aim to minimize the number of target center’s  slides required to train a quality satisfying CycleGAN.

To that end, we build various datasets each containing 1000 patches extracted from a various number $N$ of slides, train CycleGANs on each dataset and assess its performance. 
Specifically, patch datasets are extracted from $N = \{2, 5, 10, 25, 50,75\}$ slides randomly picked from target center training slides sets. 
However, we have set a fixed number of $N = 75$ slides for the source center, as our focus is specifically on investigating the impact of varying slide amounts in the target centers.
For each of these configurations, three datasets are built in total to ensure that observed results are not dependent on the chosen set of slides.

Resulting CycleGANs are once again compared through MDS1 comparison.
Finally, to ensure the results are independent of the choice of training centers, the experiment is conducted for target centers 1, 2 and 3.

\subsection{Methods comparison} \label{subsec : experiments : methods comparison}
Approaches described in \ref{subsubsec : material & method : methods : translation model} are tested and compared on the task of invasive carcinoma classification.

For comparison matter, a baseline performance is obtain on each center using the baseline classification model (see section \ref{ subsubsec : material & methods : methods : classification model}).
 
In the MDS1 approach, patches translated from center $i$ to the reference center are fed to the baseline classification model to test generalization on center $i$; in our experiments $i$ takes alternatively the values $1,2,3$.

Regarding UDA approach, the classifier is trained on both data from the reference center and data translated from the source center to centers $i,j$. The resulting classifier generalization ability is tested on center $k$ as well as on its training centers $i,j$; in our experiments $(i,j,k)$ are alternatively set to $(1,2,3)$, $(3,1,2)$, $(3,2,1)$.

\section{Results} \label{sec : Results}

\subsection{CycleGANs training stopping criterion} \label{subsec : results : cyclegans training stopping criterion}

The performance of the models trained with and without the FID stopping criterion are illustrated in the following figure \ref{fig:cyclegan_training_criterion}.
These metrics are computed using MDS1 modality.

\begin{figure}[h]
\centering
\includegraphics[width=\textwidth]{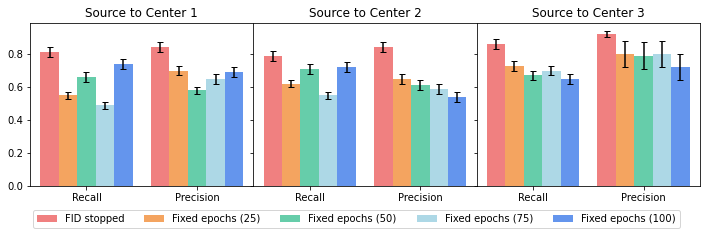}
\caption{\label{fig:cyclegan_training_criterion}Performance of CycleGAN models trained with and without the FID stopping criterion. Models are tested and compared with MDS1 comparison : by translating a center $i$ test set to the reference center's style and then evaluating the baseline classification model performance.}
\end{figure}

The stopping epochs of the FID stopping criterion for each CycleGANs is given in the following table. 

\begin{table}[h]
\centering
\begin{tabular}{cccc}
\toprule
& Source to Center 1 & Source to Center 2 & Source to Center 3 \\
\midrule
FID stopping epoch & $42 \pm 3$ & $52 \pm 9$ & $49 \pm 3$ \\
\bottomrule
\end{tabular}
\caption{\label{tab:fid_stopping_criterion_epochs}Stopping epochs and their standard deviation for CycleGAN from source to center $i$.}
\end{table}

We observe that the models trained with the FID stopping criterion consistently outperform models trained for any fixed number of epochs in terms of MDS1 performance and this, independently from the choice of the target center. For instance, for the CycleGAN\_1, the FID stopped model achieve a recall of $0.81 \pm 0.05$, and precision of $0.84 \pm 0.02$. Despite having similar or even more visually appealing generated samples, the models trained for 25 epochs perform significantly worse, with a recall of $0.55 \pm 0.10$ and precision of $0.70 \pm 0.08$.
Models' performance when trained for 50, 75 and 100 epochs are again worse with ($0.66 \pm 0.11$, $0.58 \pm 0.07$), ($0.49 \pm 0.09$, $0.65 \pm 0.06$) and ($0.74 \pm 0.10$, $0.69 \pm 0.08$) for recall and precision respectively.
Likewise, similar results are observed for CycleGANs\_2 and CycleGANs\_3 whose results are detailed in the Table \ref{tab:FID_early_stop_table} available in the \ref{appendix : results : CycleGANs training stopping criterion}.

Given the superiority of this method for training and selecting CycleGAN models, we will use the FID criterion throughout the experiments in this work.

\subsection{Slides Requirements} \label{subsec : results : slide requirements}
Regarding the number of CycleGANs' training patches, we found that using 1000 patches is sufficient. While using 10,000 patches does yield a slightly lower FID, the extended training time involved does not justify this improvement.
This is illustrated by Table \ref{tab:1000_patches}.

\begin{table}[h]
\centering
\begin{tabular}{cccc}
\toprule
 & Source to Center 1 & Source to Center 2 & Source to Center 3 \\
\midrule
100 patches & 310 / 3h & 270 / 4h  & 208 / 3h \\
1000 patches & 110 / 11h & 103 / 7h & 98 / 12h \\
10000 patches & 107 / 24h & 102 / 23h & 92 / 28h \\
\bottomrule
\end{tabular}
\caption{\label{tab:1000_patches}FID / Training time values of CycleGANs trained with 100, 1000, and 10000 patches.}
\end{table}

\begin{figure}[h]
\adjustbox{max width=1\textwidth}{
\centering
\includegraphics{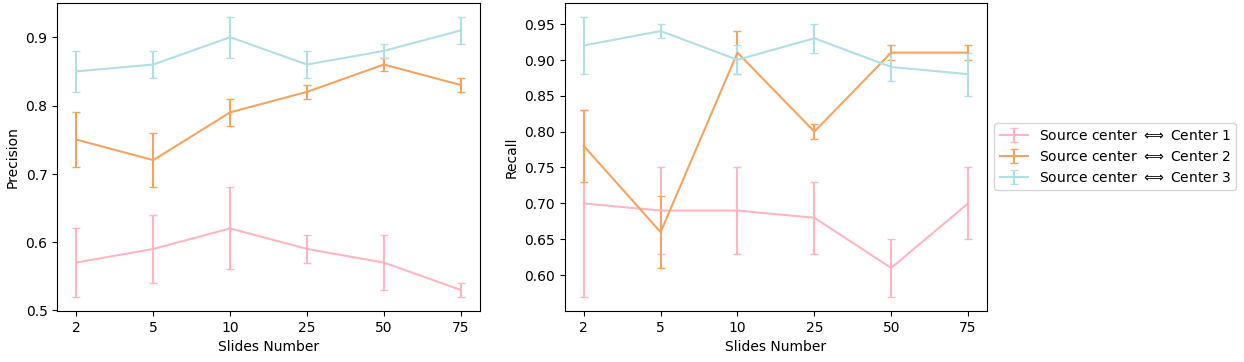}}
\caption{Performance of CycleGAN models regarding centers and slide number}
\label{fig:nb_slides_metrics_MDS1}
\end{figure}

The result of the experiment described in \ref{subsec : experiments : slides_number} regarding slides requirement are displayed in Figure \ref{fig:nb_slides_metrics_MDS1}. 

For CycleGAN\_1 and CycleGAN\_3, metrics are stable from using 2 slides to 75 slides. However, for CycleGAN\_2, using less than 10 slides negatively impacts performance.
Overall, the results indicate that the amount of slides from which training patches are extracted impacts CycleGANs performance in a center-dependent manner. 
We posit that this center dependence might be related to the amount of intra-stain variability that can be found in the center’s slides. 
Note that the results and conclusion drawn from this experiment should not be extrapolated to significantly different digital histopathology tasks without further studies.
This will be explored and discussed in detail in the following section.

\subsection{Methods comparison} \label{subsec : results : methods comparison}
In this section, we compare the different methods presented in section \ref{subsubsec : material & method : methods : translation model} following the experiment scheme described in section \ref{subsec : experiments : methods comparison}. Results are described in Table \ref{tab:methods_results}.

\begin{table}[h]
\centering
\begin{tabular}{cccccc}
\toprule
& & Source center & Center 1 & Center 2 & Center 3 \\
\midrule
\multirow{3}{*}{Baseline} & precision & 93\% & 48\% & 63\% & 75\% \\
& recall & 80\% & 39\% & 86\% & 91\% \\
& AUC & \textbf{0.98} & 0.57 & 0.75 & 0.84 \\
\midrule
\multirow{3}{*}{MDS1} & precision & - & 60\% & 86\% & 94\% \\
& recall & - & 78\% & 94\% & 94\% \\
& AUC & - & 0.81 & 0.94 & 0.91 \\
\midrule
\multirow{3}{*}{MDS2} & precision & - & 90\% & 46\% & 94\% \\
& recall & - & 80\% & 86\% & 94\% \\
& AUC & - & \textbf{0.93} & 0.88 & 0.96 \\
\midrule
\multirow{3}{*}{UDA} & precision & 92\% & 89\% & 92\% & 97\% \\
& recall & 84\% & 89\% & 95\% & 93\% \\
& AUC & 0.95 & 0.92 & \textbf{0.98} & \textbf{0.98} \\
\bottomrule
\end{tabular}
\caption{\label{tab:methods_results}Quantitative results of the different methods. UDA model is trained on samples from both reference center and augmented samples from center (1, 2) and its generalization ability is tested on center 3.}
\end{table}

\textbf{Baseline.} The baseline classification model achieved 93\% precision and 80\% recall on the reference center.
On centers 1, 2 and 3, precision and recall are 48\%, 39\%; 63\%, 86\% and 75\%, 91\% respectively. These results show that the performance of the baseline classification model on centers other than the one on which it was trained is significantly hampered. From these results, it can be argued that the generalization task is harder for some centers, as illustrated by the model's abysmal performance on center 1.

\textbf{MDS1.} On centers 2 and 3, the performance of MDS1 was significantly better than the baseline, with precision and recall being 86\%, 94\%; 94\%, 94\% respectively. The performance of MDS1 on center 1 was lower overall, with 60\% precision and 78\% recall. However, the baseline performance on this center was already significantly lower compared to the other centers. 
 
\textbf{MDS2.} In this case, metrics are also substantially better than the baseline on most of centers. Center 3 results are equal to those of MDS1, but on center 1, the model has 90\% precision and 80\% recall which is significantly better than previous method.
Results on center 2 contrast with that of other centers as precision is only 46\%.

\textbf{UDA.} UDA method shows consistent performance across centers. 
For center 1, 2 and 3, precision and recall are 96\%, 70\%, 86\%, 95\% and 97\%, 93\% respectively.
Compared to the baseline, generalization performance is significantly higher and is often even higher than baseline performance on the reference center.
This method outperform the others in term of performance and generalization ability.

Moreover, UDA performance is not impacted by the choice of training and unknwown centers, as illustrated by the following Table \ref{tab:UDA_center_impact}.

\begin{table}[ht]
\centering
\begin{tabular}{ccccccc}
\toprule
Augmentation centers & Test centers & Metrics & Source center & Center 1 & Center 2 & Center 3 \\
\midrule
 & & precision & 92\% & 89\% & 92\% & 97\% \\
(1,2) & 3 & recall & 84\% & 89\%  & 95\% & 93\% \\
& & AUC & \textbf{0.95} & 0.92 & \textbf{0.98} & \textbf{0.98} \\
\midrule
 & & precision & 93\% & 90\%  & 93\% & 98\% \\
(3,1) & 2 & recall & 80\% & 85\% & 88\% & 93\% \\
& & AUC & 0.88 & \textbf{0.96} & 0.90 & 0.93 \\
\midrule
 & & precision & 91\% & 90\% & 92\% & 98\% \\
(3,2) & 1 & recall & 84\% & 85\% & 95\% & 93\% \\
& & AUC & 0.85 & 0.93 & 0.89 & 0.94 \\
\bottomrule
\end{tabular}
\caption{\label{tab:UDA_center_impact}UDA method results depending on the choice of training and unknown centers. UDA model is trained on samples from both reference center and augmented samples from center $i,j$ and its generalization ability is tested on unknown center $k$, where $(i,j,k)$ are alternately set to $(1,2,3)$, $(3,1,2)$ and $(3,2,1)$}
\end{table}

The previous results stem from a patch-level analysis. We also computed baseline and UDA method's metrics at a slide-level to verify its validity in such a case by running the algorithm on every slide from test sets described in Section \ref{subsec : material & methods : datasets}. Precisely, the methods are still applied at a patch-level, but the metrics are globally computed on slides. For ground-truthing and metric computation, a slide is considered to be of class IC if at least 10\% of its patches are classified as IC. 
For this analysis, we used a model trained on reference center and augmented samples from every center. These results are further illustrated in Table \ref{tab:UDA_slide_analysis}

\begin{table}[h]
\centering
\begin{tabular}{cccccc}
\toprule
& Metrics & Source center & Center 1 & Center 2 & Center 3 \\
\midrule
 & precision & 98\% & 20\% & 27\% & 23\% \\
Baseline & recall & 93\% & 25\% & 40\% & 19\% \\
& F1-score & \textbf{96}\% & 19\% & 17\% & 21\% \\
\midrule
& precision & 98\% & 97\% & 97\% & 97\% \\
UDA & recall & 95\% & 95\% & 94\% & 95\% \\
& F1-score & \textbf{96}\% & \textbf{97}\% & \textbf{97}\% & \textbf{97}\% \\
\bottomrule
\end{tabular}
\caption{\label{tab:UDA_slide_analysis}UDA method's slides analysis. 
UDA model is trained on samples from both reference center and augmented samples from center (1, 2, 3).}
\end{table}

\section{Discussion.} \label{sec : Discussion}
\textbf{CycleGANs training stopping criterion}.
Overall, utilizing FID stopping criterion to end CycleGANs training consistently led performance improvement compared to stopping at a arbitrarily fixed epoch.
Furthermore, FID stopped models had significantly lower error margin, indicating that the method is consistent and produces models that perform similarly. 
Such discrepancy in fixed training epochs is yet another proof of CycleGANs training instability and highlights the importance of utilizing more robust and automatic stopping criteria such as the one used in this work. 
Additionally, using the FID criterion resulted in significant time savings as training often stopped in the 25-40 epochs range. In our settings, this reduced training time by approximately 10 hours compared to the model trained for 100 epochs.

On the other hand, it should be noted that this method may not always produce the optimal CycleGAN model for the task at hand. 
This would require an evaluation of models at each epoch during training, which is a tedious and costly process. 
However, it is useful to produce performing enough models for our task at hand in an automated fashion.
Consequently, this makes it quite easy to include in a more systematic and production-oriented workflow when working with new and exotic stain variations.

\textbf{Slides requirements}.
Section \ref{subsec : results : slide requirements} indicates that the amount of slides impacts results for one out of three centers. This clearly shows that this impact is center dependent. 
We hypothesize that this center dependence is related to the amount of intra-stain variability that can be found in a particular center’s slides. 
Hence, centers showing low intra-stain variability may require only 2 slides for the cycleGAN to capture this diversity, whereas in the case of high intra-stain variability, it may not be sufficient.

To validate this hypothesis, patches from $x$ slides of each center are converted to the Hue-Saturation-Value color (HSV) channel, Hue's mean and its standard deviation is computed for every patch. 
Here, $x$ takes the values $\{2,5,10,25,50,75)\}$.
The results are displayed in the following figure \ref{fig:results_intra_stain}. 

\begin{figure}[ht]
\adjustbox{max width=\textwidth}{
\centering
\includegraphics{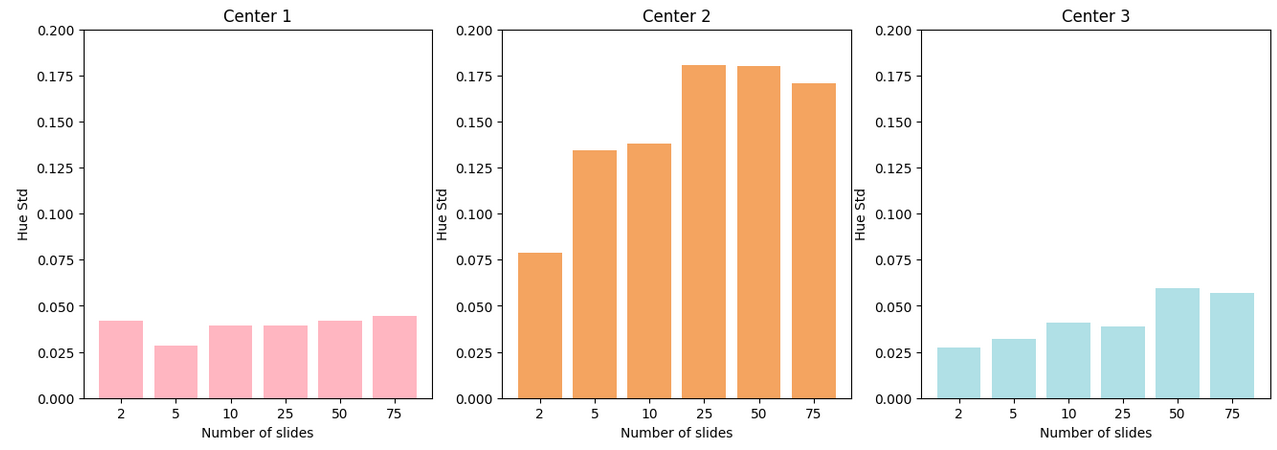}}
\caption{Hue standard deviation by center and slides number.}
\label{fig:results_intra_stain}
\end{figure}

The results indeed show a clear increase in hue’s standard deviation when increasing the number of slides for center 2, from 0.076 with 2 slides to a stable 0.178 when hitting the 25 slides mark, whereas it stays low and stable in the case of center 1 and 3.
The evidence suggests that some centers exhibit higher intra-stain variability, and the amount of slides required to train good stain translators depends on this prior. 
Besides, it also suggests that building more refined tools capable of assessing this stain variability in a straight-forward and scientific manner could be an interesting and useful work to better assess cycleGANs training data requirements.

\textbf{Methods comparison}. Overall, these results indicate that every method outperform baseline performance.
Nevertheless, it seems that the performance gains are center dependent and are also impacted by translation direction.
This indicate that some stains translations are more challenging for the CycleGAN to learn. 
There seems to be an emergent "translation direction" effect when comparing MDS1 and MDS2 results for center 1 and 2.
This translation's direction impact on performance has been first observed in this paper \cite{gadermayr_MDS}, whose advice is to translate from the "hard-to-segment" to the "easy-to-segment" stain. However, in their work, the task on top of the generation process was glomeruli segmentation and translation were made between a Periodic-Acid-Sciff (PAS) stain and IHC stains, in particular CD31.
In this case, "hard" and "easy" to segment domains are quite easy to determine, as PAS will make the glomeruli structure obvious while CD31 will focus on targeting lymphocytes. Hence, one would focus on MDS1 for successful segmentation, and indeed, the authors found it to be the "preferred method".
In the present work however, every stain is H\&E and H\&E\&S and typically highlight the same structures of interest and in fact, no difference of performance has been observed between H\&E and H\&E\&S during this study. Hence, "hard" and "easy" to classify stains don't hold the same meaning and cannot be predicted beforehand.
This translation direction effect might actually be explained by the high intra-center stain diversity observed in center 2's samples and discussed in section \ref{subsec : results : slide requirements}.
To elaborate, a plausible explanation might be that the CycleGAN focused on an "easy-to-translate" variation of center 2 stain, which would guarantee a low cycle-loss, but is not representative of the entire real distribution. Plus, this hypothesis also explains why this problem doesn't occur on the target to source translation, as source stain doesn't show such intra-center stain diversity. 
This behaviour can be seen as a clear case of mode collapse for CycleGANs and deserves a further investigation.

Next, on the methods themselves, each has it's specificity and use-case. 
MDS1 doesn't require any classifier re-training. However, if one's aim is to compute metrics of any kind, this method requires labeled data from the target center.
On the other hand, MDS2 transfers the labelling knowledge on a source center to the target center as labels are preserved with the translation. Thus, it is possible to use a reference center's dataset to train a classifier on another domain and thus, no labeling on this domain is needed.
These two methods are stain specific as obtained classifiers can only be used with one unique stain at the end of the process. This is the main limitation of these methods as they can suffer from center-wise difficulty and high intra-center stain diversity as explained previously.

Finally, UDA method comes with two merits.
First, this method yields the best and most consistent results overall. Furthermore, the model obtained with this method can be used on several stains and even on unseen ones, which is a great advantage from a practical point of view. We posit that the greater the number of stains available for training, the greater the potential for generalization.
Undoubtedly, UDA method is the most convenient and preferred method for high performance and stain robustness purposes.

\section{Conclusion} \label{sec : Conclusion}
The hereby study is built around three experiments that each bring an answer to major issues in digital pathology.
Firstly, it introduces a systematic and consistent method to train and choose a CycleGAN model using FID metric, thus answering a quite often discussed concern regarding CycleGAN usage in a medical context.
The study also explores a relatively unconsidered but nonetheless decisive topic regarding digital pathology, that is, the CycleGANs' training WSI requirement. Basically, as the intra-center (i.e intra-scanner) stain diversity increases, so does the WSI requirement for training the CycleGAN.
Finally and most importantly, CycleGANs appear to be a solution to the generalization challenge in digital histopathology. 
The study shows improved breast carcinoma classification performance when using all three CycleGAN's methods on every stains we had at our disposal compared to the baseline. Each method has its own use-case and merits and can answer any stain related domain adaptation problem.

To sum up, we deliver a CycleGAN based solution to the generalization challenges in digital histopathology.

The method have demonstrated successful application in the specific task of breast invasive carcinoma classification. Further studies are to be performed to confirm the effectiveness of this method on various histopathology tasks.
Another limitation of this study is to consider the staining and the scanning protocol as one entity. More precisely, the variability observed in this study might stem as much from the staining protocol than the scanning process, but in our case, no difference has been made between these sources of variability. A more precise approach would consider how each step independently is accounted for by our methodology. A further study is required to explore this possibility.
Moreover, future work may focus on tweaking generator and discriminator architecture and hyper-parameters, as this topic was left aside during this study although it may improve performance compared to the actual work. Another interesting area of research would be to investigate the benefits of using this method with the most recent generative models capable of image-to-image tasks compared to the conventional CycleGAN architecture.

\newpage
\appendix

\section{Results} \label{appendix : results}
\subsection{CycleGANs training stopping criterion}\label{appendix : results : CycleGANs training stopping criterion}

Detailed results for CycleGAN's training stopping criterion experiment for CycleGAN\_1, CycleGANs\_2 and CycleGANs\_3. 
Results show improved perfomance when using the FID stopping criterion independently from the 
choice of the target center.

\begin{table}[h]
\centering
\begin{tabular}{cccc}
\toprule
Centers & Model & Recall & Precision \\
\midrule
 & \textbf{FID stopped ($42 \pm 3$)} & \textbf{$0.81 \pm 0.05$} & \emph{$0.84 \pm 0.02$} \\
 & Fixed epochs (25) & $0.55 \pm 0.10$ & $0.70 \pm 0.08$ \\
Source to center 1 & Fixed epochs (50) & $0.66 \pm 0.11$ & $0.58 \pm 0.07$ \\
 & Fixed epochs (75) & $0.49 \pm 0.09$ & $0.65 \pm 0.06$ \\
 & Fixed epochs (100) & $0.74 \pm 0.10$ & $0.69 \pm 0.08$ \\
\midrule
 & \textbf{FID stopped ($52 \pm 9$)} & $0.79 \pm 0.06$ & $0.84 \pm 0.04$\\
 & Fixed epochs (25) & $0.62 \pm 0.10$ & $0.65 \pm 0.07$ \\
Source to center 2 & Fixed epochs (50) & $0.71 \pm 0.10$ & $0.61 \pm 0.06$ \\
 & Fixed epochs (75) & $0.55 \pm 0.10$ & $0.59 \pm 0.08$ \\
 & Fixed epochs (100) & $0.72 \pm 0.11$ & $0.54 \pm 0.06$ \\
\midrule
 & \textbf{FID stopped ($49 \pm 3$)} & $0.86 \pm 0.06$ & $0.92 \pm 0.03$ \\
 & Fixed epochs (25) & $0.73 \pm 0.10$ & $0.80 \pm 0.07$ \\
Source to center 3 & Fixed epochs (50) & $0.67 \pm 0.11$ & $0.79 \pm 0.08$ \\
 & Fixed epochs (75) & $0.70 \pm 0.10$ & $0.80 \pm 0.08$ \\
 & Fixed epochs (100) & $0.65 \pm 0.10$ & $0.72 \pm 0.08$ \\
\bottomrule
\end{tabular}
\caption{\label{tab:FID_early_stop_table}Performance of CycleGAN models trained with and without the FID stopping criterion. Utilizing FID criterion led to better performance in every case.}
\end{table}


\bibliographystyle{spiejour}   

\listoffigures
\listoftables

\end{spacing}
\end{document}